Critical temperatures of two- and three-dimensional Ferromagnetic Ising models- A hierarchy

M V Vismaya and M V Sangaranarayanan*

Department of Chemistry

Indian Institute of Technology Madras-Chennai-600036 India

E mail: vismayaviswanathan11@gmail.com; E mail: sangara@iitm.ac.in

Abstract

A unified algebraic structure is shown to exist among various equations for the critical temperatures pertaining to diverse two- and three-dimensional lattices. This isomorphism is a pointer to the straight-forward extension of two-dimensional results to corresponding three dimensional analogues.

1.Introduction

The estimation of critical temperature of three- dimensional Ising models has been an active area of research in statistical physics. Onsager's solution of the square and rectangular lattices is a mile stone in this context on account of the exact formulation of the partition function and associated thermodynamic quantities [1]. The critical temperature for square lattices follows from the spontaneous magnetization equation given by [2]

$$M_0 = \left(1 - \frac{1}{\sinh^4\left(\frac{2J}{kT}\right)}\right)^{\frac{1}{8}} \quad (1)$$

where $J$ is the customary nearest neighbour interaction energy and k denotes the Boltzmann constant. The partition function $Q$ for $N$ sites is expressed in the thermodynamic limit as [3]

$$\left(\frac{1}{N}\right)\log Q = \log[\sqrt{2}\cosh(2K)] + \left(\frac{1}{\pi}\right)\int_0^{\pi/2} d\phi \log\left[1 + \sqrt{1 - \kappa^2 \sin^2(\phi)}\right] \quad (2)$$

with K = J/kT. The numerical integration in equation (2) is straight-forward; however, the integral can also be evaluated exactly in terms of the Gauss hypergeometric function, by a simple substitution method [4]. The partition function contains not the spontaneous magnetization but a 'dimensionless interaction' parameter $\kappa$ defined as

$$\kappa = \frac{2\sinh(2K)}{\cosh^2(2K)} \quad (3)$$

We note that the critical temperature $T_c$ occurs at $\kappa = 1$ viz $\frac{J}{kT_c}$ is nearly equal to 0.44068. In this Brief Communication, we show that by merely changing the above definition of $\kappa$, several known critical temperatures of two- and three-dimensional lattices employing mean field approximation, Bethe approximation and computer simulations, can be accurately deduced. This in turn may lead to new approaches for analysis of three-dimensional Ising models, in particular for the spontaneous magnetization and zero-field partition function.

## 2. Critical temperatures of two- and three-dimensional lattices

The methodology adopted by Onsager in the analysis of two-dimensional models has led to the prediction of critical temperature at $\kappa = 1$. It is of interest to investigate whether the critical temperature can be accurately predicted by altering this definition of $\kappa$ for other two and three-dimensional models. Intuitively, one expects that either the interaction term '$K$' or numerical factor '2' in equation (3) may get altered while analyzing the corresponding three-dimensional models, with appropriate consideration of the coordination numbers.

Table1: Equations for the critical temperature ($T_c$) for various two- and three-dimensional lattices

| Model | Equation for $\kappa$ in terms of $K = \frac{J}{kT}$ | $\frac{J}{kT_c}$ |
|---|---|---|
| 2-D Ising (mean field approximation) | $\kappa = 4K = 1$ | 0.25 |
| 2-D Ising (Bethe approximation) | $\kappa = 3\tanh(K) = 1$ | 0.346573 |
| 2-D Ising (Onsager's exact solution) | $\kappa = \dfrac{2\sinh(2K)}{\cosh^2(2K)} = 1$ | 0.4406867 |
| 2-D Ising (exact) honeycomb lattice | $\kappa = \dfrac{1}{2}\cosh(2K) = 1$ | 0.658478 |
| 2-D Ising (exact) triangular lattice | $\kappa = \sqrt{3}\sinh(2K) = 1$ | 0.274653 |
| Kagome Lattice | $\kappa = \dfrac{e^{4K}}{3 + 2\sqrt{3}} = 1$ | 0.466566 |
| 3-D Ising (simple cubic) | $\kappa = \dfrac{2\sinh(1.9289\,K)}{\sinh^{-1}(1)} = 1$ | 0.22165 |
| 3-D Ising (face centred cubic) | $\kappa = 5\tanh(2K) = 1$ | 0.101366 |
| 3-D Ising (body -centred cubic) | $\kappa = \dfrac{\tanh(6K)}{\sqrt{3} - 1} = 1$ | 0.155522 |
| 2-D Ising (honeycomb lattice) Bethe approximation | $\kappa = 2\tanh(K) = 1$ | 0.549306 |
| 3-D Ising (diamond lattice) Bethe approximation | $\kappa = \dfrac{4}{3}\sinh(2K) = 1$ | 0.346573 |
| 3-D Ising (triangular lattice) Bethe approximation | $\kappa = 5\tanh(K) = 1$ | 0.202732 |
| 3-D Ising (face centred cubic) Bethe approximation | $\kappa = 2\tanh(6K) = 1$ | 0.091551 |
| 3-D Ising (body -centred cubic) Bethe approximation | $\kappa = \dfrac{2\sinh(6K)}{\cosh^2(6K)} = 1$ | 0.146895 |
| 3-D Ising (diamond lattice) | $\kappa = \dfrac{\sinh(2.15\,K)}{\sinh^{-1}(1)} = 1$ | 0.3697477 |

Table 1 provides the equations for critical temperatures of various lattices in two and three dimensions that may yield a unified perspective on the analysis of critical phenomena. The data on the critical temperatures have been sourced from [5,6,7]. We point out that the mathematical functions that appear in Table 1 are not unique. While alternate equations are indeed possible, our emphasis is to keep the isomorphism with Onsager's definition of κ for square lattices. While a few equations of Table 1 pertaining to two-dimensional systems are already known either through star-triangle transformation or by mapping of series expansions at low and high temperatures, three entries of Table 1 are of much interest.

(a) Critical temperature from mean field approximation

It is well-known that the mean field approximation neglects local correlations and the equation for the critical temperature is $\frac{J}{kT_c} = \frac{1}{4}$ for square lattices [6] with J being the customary nearest neighbour interaction energy. At the critical temperature, the parameter $\kappa = 4K = 1$. If the hyperbolic terms of equation (3) are expanded using exponentials and truncated to the linear term in $K$, it follows that $4K = 1$ at the critical temperature, which is the mean field prediction for square lattices. This reduction of the qualitatively incorrect mean field result by approximation of the Onsager's exact solution is un-anticipated.

(b) Critical temperature using Bethe approximation

The well-known Bethe approximation [7] which introduces 'local order' and provides a better approximation for $T_c$ than the mean field approximation also seems to follow from

$$\kappa = 3\,tanh(K) = 1 \tag{4}$$

yielding $K$ at the critical temperature as

$$\frac{J}{kT_c} = 0.346573 \tag{5}$$

for square lattices and is in exact agreement with [7]. For triangular, honeycomb and Kagome lattices, the equations for $\kappa$ are shown in Table 1.

(c) Critical temperature of three-dimensional lattices using computer simulation

A major un-solved problem in the statistical physics of lattice models is the construction of exact solution of three-dimensional Ising models, even at zero magnetic fields. However, accurate estimates of the critical temperatures are known in three dimensions, for various cubic lattices using extensive computer simulations. The consensus is that the equation for the critical temperature is given by $\frac{J}{kT_c} = 0.221654$ using computer simulations in conjunction with series expansions [9] for simple cubic lattices. As shown in [10], this exact value is often a touchstone for validating any perceived exact solution of three-dimensional systems.

It is logical to expect that appropriate modification of equation (3) should yield accurate critical temperatures for different three-dimensional lattices. In the case of simple cubic, face-centered cubic, body-centered cubic and diamond lattices in three dimensions, a simple alteration of equation (3) is shown to yield the critical temperatures which are in excellent agreement with the literature estimates:

$$\kappa = 5\tanh(2K) = 1 \quad \text{(face-centered cubic)} \tag{6}$$

$$\kappa = \frac{2\sinh(1.9289\,K)}{\sinh^{-1}(1)} = 1 \text{ (simple cubic)} \tag{7}$$

$$\kappa = \frac{\tanh(6K)}{\sqrt{3}-1} = 1 \quad \text{(body-centered cubic)} \tag{8}$$

$$\kappa = \frac{\sinh(2.15\,K)}{\sinh^{-1}(1)} = 1 \quad \text{(diamond lattice)} \tag{9}$$

All the above equations for various three-dimensional lattices have solely resulted from altering the argument of the hyperbolic terms in the original equation for $\kappa$ due to Onsager [1], for square lattices. This important inference indicates that the exact solution of three-dimensional Ising-like models should involve re-formulation of $\kappa$ as shown in Table 1 in the equations for the partition function, spontaneous magnetization, internal energy, entropy and specific heat. It is also of interest to note that all equations (except the mean field ones) consist of exponential or trigonometric or hyperbolic functions. Hence the field-dependent magnetization equations of various two and three-dimensional lattices too will possess similar equations *albeit* with the corresponding critical exponents.

3.Summary

The critical temperatures of various two- and three-dimensional lattices are shown to have a common simple algebraic form involving essentially minor changes in the interaction energy terms of Onsager's exact solution. For various lattices analyzed using Bethe approximations or using computer simulations, the values arising from the equations postulated here are in excellent agreement with the reported values.

Acknowledgements

This work was supported by the Mathematical Research Impact Centric Scheme (MATRICS) of SERB, Government of India. We thank the Computer Centre IIT Madras for computational resources.